\documentclass[prl,aps,amssymb,showpacs,twocolumn]{revtex4}
\usepackage{amsmath}
\usepackage{amssymb}
\usepackage{amsthm}
\usepackage{amsfonts}
\usepackage{enumerate}
\usepackage{latexsym}
\usepackage{bm}

\newcommand{\ben}{\begin{enumerate}}
\newcommand{\een}{\end{enumerate}}
\newcommand{\beq}{\begin{equation}}
\newcommand{\eeq}{\end{equation}}
\newcommand{\bse}{\begin{subequation}}
\newcommand{\ese}{\end{subequation}}
\newcommand{\bea}{\begin{eqnarray}}
\newcommand{\eea}{\end{eqnarray}}
\newcommand{\bc}{\begin{center}}
\newcommand{\ec}{\end{center}}

\newcommand{\bp}{\mbox{\boldmath $\pi$}}
\newcommand{\ba}{\mbox{\boldmath $\alpha$}}
\newcommand{\bb}{\mbox{\boldmath $\beta$}}
\newcommand{\be}{\mbox{\boldmath $\eta$}}
\newcommand{\bt}{\mbox{\boldmath $\theta$}}
\newcommand{\bvep}{\mbox{\boldmath $\varepsilon$}}
\def\r{r_0}

\begin{document}

\title{\Large\bf Noncommutative Coulombic Monopole}

\author{\bf Stefano Bellucci$^1$ and Armen Yeranyan$^2$}

\affiliation{\it $^1$ INFN-Laboratori Nazionali di Frascati,
 P.O. Box 13, I-00044, Frascati, Italy\\
$^2$ Yerevan State University, Alex  Manoogian St., 1, Yerevan,
375025, Armenia}

\begin{abstract}

We have constructed the appropriate Hamiltonian of the noncommutative coulombic monopole (i.e. the noncommutative hydrogen atom with a monopole).
The energy levels of this system have been calculated, discussed and compared with the noncommutative hydrogen atom ones. The main emphasis is put
on the ground state.
In addition, the Stark effect for the noncommutative coulombic monopole has been studied.

\end{abstract}


\maketitle

\setcounter{equation}0
\noindent

\section{Introduction}

Recently many papers have been devoted to the construction and the study of noncommutative systems.
The reasons of this interest were the predictions of String theory \cite{string} in the low-energy limit, which along with
the Brane-world scenario \cite{bran}, led to the fact that space-time could be noncommutative.
On the other hand, a considerable attention was paid to the investigation of the phenomenological consequences of noncommutativity.
The noncommutative deformation of the Standard model  was suggested \cite{standart} (for the
renormalization of the energy-momentum tensor in noncommutative field theories, see
\cite{bbk}).
The  Quantum Hall effect on noncommutative spaces was also studied \cite{hall}.
For these purposes noncommutative quantum mechanics was also extensively studied \cite{ncqm}.
There, careful investigations showed that it will be hard to connect noncommutativity with reality on the experimental level.
This is due to the fact that noncommutativity practically plays no role in most "attraction" problems \cite{ho},
and in scattering problems it becomes essential only for the case when the potential does not rapidly converge \cite{scat}.
However, in spite of all of this, there are several phenomena which allow us to hope for getting a connection between noncommutativity and experiments.
One of these phenomena
can be due to the presence of a magnetic field in noncommutative systems.
As it was shown "turning on" a magnetic field in a noncommutative system can lead to such
interesting phenomena as the appearance of two phases separated by a critical point, etc. \cite{magn}.
Also it was suggested a method, based on the presence of a magnetic field in a system (Rydberg atom),
which gives us the possibility to test spatial noncommutativity \cite{zhang}.

Indeed, in the noncommutative case only systems in the presence a constant magnetic field were more or less thoroughly investigated.
However, for the commutative case, not only such systems, but also quantum-mechanical systems with a
Dirac monopole were studied (see for example \cite{val} and references therein).
This is due to the rich symmetries of the magnetic field of the monopole.
From this point of view it is very useful to study noncommutative quantum-mechanical systems in the presence of a Dirac monopole.
This allows us to understand more about the nature of both noncommutativity and the monopole.

Here we consider a noncommutative coulombic monopole ($NCM$), i.e. a noncommutative MIC-Kepler system.
The first abortive attempt to the consideration of this system was made in \cite{denis}.
The investigation of the system is interesting not only owing to the simultaneous presence of magnetic and "noncommutativity" fields,
but also for the reason that
the commutative coulombic monopole ($CCM$) is the system most similar to the hydrogen atom.

The $CCM$ was constructed by Zwanzinger and later
rediscovered by McIntosh and Cisneros  \cite{Z}. The $CCM$ describes the charge motion
in the field of the Dirac dyon and its Hamiltonian has the form
\bea \hat{H}^{(CCM)}=\frac{1}{2M}\hat{\tilde{{\bf p}}}^2
+\frac{{\hbar}^2{s}^2}{2 M r^2}-\frac{\gamma}{r},
 \label{Ham1} \eea
where momenta and coordinates satisfy the following commutation relations:
\bea
\left[\hat{\tilde{p}}_i,\hat{\tilde{p}}_j\right]=\imath \hbar^2 s\; \varepsilon_{ijk} \frac{x^k}{r^3}, \quad
\left[x^i,\hat{\tilde{p}}_j\right]=\imath \hbar \delta^i_j.
\eea
The momentum $\hat{\tilde{{\bf p}}} $ is connected with the "canonical" one as follows:
\bea
\hat{\tilde{{\bf p}}}=\hat{{\bf p}}- {\bf A}(\bf r),
\eea
where ${\bf A}(\bf r)=\hbar s\frac{{\bf r}\times{\be}}{r(r-{\bf r}\cdot{\be})}$ is the vector-potential of the monopole field
and ${\be}$ is a unit constant vector with arbitrary direction. Below we will take it along the $z$ axis.

The main reason which allows us to consider $CCM$ as generalization of the hydrogen atom is the hidden symmetry given by the following constants of motion:
\bea \hat{{\bf J}}={\bf r}\times
\hat{\tilde{{\bf p}}} -\hbar s\frac{{\bf r}}{r}, \quad
\hat{{\bf I}}=\frac{1}{2M}\left[\hat{{\bf J}}\times \hat{\tilde{{\bf p}}}
-\hat{\tilde{{\bf p}}}\times{\hat{ {\bf J}}}\right] -\gamma\frac{{ {\bf r}}}{r}. \label{constmot}
\eea
These constants of motion, together with the Hamiltonian, form the
quadratic symmetry algebra of the Coulomb problem.
The operator  $\hat{{\bf J}}$ defines  the angular momentum of the
system, while the operator $\hat{{\bf I}}$ is  the analog of the Runge-Lenz
vector.
For fixed negative values of the  energy,  the constants of motion
form  the $so(4)$ algebra.  Let us mention that the
monopole number $s$ satisfies the Dirac's  charge
quantization rule,  $s=0,\pm1/2,\pm 1,\ldots$.

This Letter is organized as follows.
In the next section the noncommutative Hamiltonian for the hydrogen atom in the presence of a monopole is constructed.
Afterwards, the energy correction due to noncommutativity is found and discussed. In the last section the Stark effect in this system  is studied.
In the Conclusion the main results are summarized.

\section{The symplectic form and the construction of the noncommutative Hamiltonian}

Let us have a commutative system with an arbitrary magnetic field. We would like to construct the corresponding noncommutative system. For this
purpose we must somehow introduce noncommutativity (not necessary constant) in our system. This can be done by specifying the Hilbert space
and the phase space. The Hilbert space can be taken to be exactly the same as the Hilbert space of the corresponding commutative system \cite{chain}.
As for the phase space, it must be changed, in order to include noncommutativity. The only condition we impose, is that this change must not violate
(at least in a general sense, see below) the Jacobi identities.
This becomes really essential in the case of non-constant magnetic and/or "noncommutativity" fields.
The easiest correct way to "turn on" noncommutativity is writing down an appropriate Ansatz for the symplectic structure.

Let us define a quite general symplectic structure as follows:
\bea
\Omega&=&d\pi_i \wedge dq^i+ \nonumber \\
&+&\frac{1}{2} \varepsilon_{kij} B^k({ \bf q}) \;dq^i \wedge dq^j+
\frac{1}{2} \varepsilon^{kij} \Theta_k(\bp) \; \pi_i \wedge \pi_j \label{symps}.
\eea
Here (and below) the indexes $i,\,j,\,k$ run over values $(1,\,2,\,3)$.
The ${\bf B}({\bf q})$ and ${\bf \Theta}({\bp})$ are vectors of magnetic and "noncommutativity" fields accordingly.
As it follows from the closure of the symplectic structure (\ref{symps}), they have to satisfy the condition
\beq
div {\bf B}=div{\bf \Theta}=0.
\eeq
Let us mention that, in the case of a monopole field, the first condition is violated at the origin \cite{jack}.
The commutation relations corresponding to the symplectic form $\Omega$ have the form
\bea
\left[\hat{q}^i,\hat{q}^j\right]=\imath \hbar && \frac{\varepsilon^{ijk} \Theta_k}{1-B^n \Theta_n}, \quad
\left[\hat{\pi}_i,\hat{\pi}_j\right]=\imath \hbar \frac{\varepsilon_{ijk} B^k}{1-B^n \Theta_n}, \nonumber \\
& &\left[\hat{q}^i,\hat{\pi}_j\right]=\imath \hbar \frac{\delta^i_j-B^i \Theta_j}{1-B^n \Theta_n}.
\eea
We will consider only the case ${\bf \Theta}\ll1$ and, in all expressions below, we will keep only the zeroth and the first order terms in ${\bf \Theta}$.
Now let us bring the commutation relations to the canonical form. We can manage to do this by introducing
new variables which are connected with the old ones as follows:
\bea
x^i=\hat{q}^i-\beta^i\left(\hat{{\bp}}\right),
\; \hat{p}_i=\hat{\pi}_i+\alpha_i\left(\hat{{\bf q}}\right)-
 \frac{\partial \alpha_n\left(\hat{{\bf q}}\right)}{\partial \hat{q}^i} \;\beta^n\left(\hat{{\bp}}\right).
\eea
The regenerative connection has the form
\bea
\hat{q}^i=x^i+\beta^i(\hat{\tilde{{\bf p}}}), \; \hat{\pi}_i=\hat{\tilde{p}}_i-
\varepsilon_{ikn}B^k({\bf r})\; \beta^n(\hat{\tilde{{\bf p}}}) \label{oldnew},
\eea
where $\hat{\tilde{{\bf p}}}=\hat{{\bf p}}-{\ba }({\bf r})$, ${\ba}$ and ${\bb}$ are vector-potentials of magnetic and "noncommutativity" fields accordingly
\beq
{\bf B}=rot{\ba}, \quad {\bf \Theta}=rot{\bb}.
\eeq

Later on, in this Letter we will choose ${\ba}({\bf r})={\bf A}({\bf r})$
 and
${\bb}(\hat{\tilde{{\bf p}}})=
\frac{1}{2}{\bt}\times\hat{\tilde{{\bf p}}}$
which correspond to the case of the magnetic field of a Dirac monopole ${\bf B}=\hbar s\frac{{\bf r}}{r^3}$ and constant noncommutativity
${\bf \Theta}={\bt}=const$.

As the Hilbert space does not change, the Hamiltonian corresponding to $NCM$ has the following form:
\bea \hat{H}^{(NCM)}=\frac{1}{2M}\hat{{\bp}}^2
+\frac{{\hbar}^2{s}^2}{2 M \hat{q}^2}-\frac{\gamma}{\hat{q}}.
 \label{Ham2} \eea
Now, substituting the relations (\ref{oldnew}) in this Hamiltonian and taking into consideration the smallness of $\theta$, we get
\bea \hat{H}^{(NCM)}=\hat{H}^{(CCM)}-\frac{1}{2r^3}\left(\gamma\; {\bt}\cdot \hat{{\bf J}}-\hbar s\; {\bt}\cdot \hat{{\bf I}}\right),
 \label{Ham3}\eea
Here $\hat{H}^{(CCM)}$ is the Hamiltonian of the usual coulombic monopole, which is defined by (\ref{Ham1}), wheras
the second term (where $\hat{{\bf J}}$ is the angular momentum) and the third one (where $\hat{{\bf I}}$ is the Runge-Lenz vector) are
corrections due to noncommutativity. In the case when the monopole field is not present ($s=0$), we get the expression of the
noncommutative hydrogen atom Hamiltonian, which was obtained in \cite{chain}. In this sense, we can say that the second term in the
correction is similar to the spin-orbit coupling, but \textit{there are no analogs (in the literature) for the third term}.

\section{Spectrum for the noncommutative hydrogen atom in the presence of monopole}

Now let us calculate a "noncommutative" correction of the spectrum of the hydrogen atom in the presence of a monopole.
Here and below we take no account of the electron spin.
As it was shown in the previous section, the first order "noncommutative" correction has the following form:
\bea \Delta\hat{H}^{(NCM)}=-\frac{1}{2r^3}\left(\gamma\; {\bt}\cdot \hat{{\bf J}}-\hbar s\; {\bt}\cdot \hat{{\bf I}}\right).
 \label{Ham4}\eea
In order to calculate this correction, one must choose the appropriate coordinate system.
Due to the hidden symmetry, the $CCM$ could be factorized in a few coordinate systems, e.g. in the spherical and parabolic ones.
Because of the presence of a preferential direction (given by ${\bt}$) in the "noncommutative" correction we will use the parabolic basis.
There is another reason to justify this choice and it will be given later.

The energy spectrum and wave functions of $CCM$ in parabolic coordinates have the following form (see for example \cite{mard}):

\begin{widetext}
\begin{eqnarray}
& & \hspace{4cm} E^{(CCM)}_n=-\frac{\gamma}{2 \r n^2},\label{enerccm} \\
& &\psi_{n_1n_2m}^{(s)}(\mu,\nu,\varphi) =
\frac{\sqrt{2}}{n^2 \,r^{3/2}_0}\;\Phi_{n_1m_1}\left(\frac{\mu}{n \r }\right)
\Phi_{n_2m_2}\left(\frac{\nu}{n \r}\right)\,\frac{e^{i(m-s)\varphi}}{\sqrt{2\pi}},
\label{parwave1} \\
& &\Phi_{n_im_i}(x) = \frac{1}{\Gamma(m_i+1)}
\sqrt{\frac{\Gamma(n_i+m_i+1)}{(n_i)!}}\,\,e^{-\frac{
x}{2}}\,\,( x)^{\frac{m_i}{2}}\,\,F(-n_i; m_i+1;
x). \label{parwave2}
\end{eqnarray}
\end{widetext}

Here $n$ is a principal quantum number, connected with the parabolic quantum numbers $n_1$ and $n_2$ as follows:
\bea
 \nonumber
n=n_1+n_2+m_+ +1 \label{pqn},
\eea
where $m_{\pm}=\frac{m_2 \pm m_1}{2}$, $m_1=|m-s|$, $m_2=|m+s|$, $m$ is the azimuthal quantum number, and $\r=\frac{\hbar^2}{\gamma M}$.
Let us mention that, in the parabolic coordinate system, the two quantities which have definite values, along with the energy,
are the projections of the angular momentum and the Runge-Lenz vector on the $z$ axis \cite{mard}
\begin{widetext}
\bea
\hat{J}_z\psi_{n_1n_2m}^{(s)}(\mu,\nu,\varphi)&=&\hbar \, m \, \psi_{n_1n_2m}^{(s)}(\mu,\nu,\varphi),\nonumber \\
\hat{I}_z\psi_{n_1n_2m}^{(s)}(\mu,\nu,\varphi)&=&-\frac{\gamma}{n} \, (n_1-n_2-m_-)\, \psi_{n_1n_2m}^{(s)}(\mu,\nu,\varphi). \label{values}
\eea
\end{widetext}
This fact allows us find the energy correction in the easiest way.
Directing the $z$ axis along ${\bt}$ we can immediately bring the correction (\ref{Ham4}) into a convenient form.
Clearly, this is the second reason justifying the choice of parabolic
coordinates.

The expression of the matrix elements of the correction has the following form:
\beq
\Delta H^{(NCM)}_{n'_1n_1\,m'm}=\langle{n'_1n'_2m'}|\Delta \hat{H}^{(NCM)}|{n_1n_2m}\rangle,
\eeq
where we suppose $n=n'$.
In order to find this matrix elements, let us direct the $z$ axis along ${\bt}$,
use the eigenvalues (\ref{values}) and take into consideration the matrix elements
\bea
\nonumber
\langle{n'_1}|\frac{1}{r^3}|{n_1}\rangle=\frac{(-1)^{n'_1+n_1}}{\r^3 n^3}\sum^{n-1}_{j=m_+}
\frac{ C^{j,m_+}_{a,\,a'_1;\; b, \,b'_1} C^{j,m_+}_{a,\,a_1;\; b, \,b_1}}{j(j+1)(j+1/2)},
\eea
where $C^{j,m_+}_{a,\,a_1;\; b, \,b_1}$ are Clebsch-Gordan coefficients,
$a=\frac{n+m_- -1}{2}$, $a_1=\frac{m_-+n-1-2n_1}{2}$, $b=\frac{n-m_- -1}{2}$, $b_1=\frac{2m_+ +1-m_- -n+2n_1}{2}$,
and primed indices are given by similar expressions, obtained with the
only replacement of $n_1$ by $n_1'$. Here we omit the index $m$, because the matrix is diagonal in $m$.
Let us notice that the expression of the matrix elements in the above
can be easily obtained using the interbasis expansion (generalized Park-Tarter formula)\cite{mard}
and the mean value of $\frac{1}{r^3}$ \cite{landau}. Hence,
for the matrix element of the correction we get
\bea
&&\Delta H^{(NCM)}_{n'_1n_1}= \label{matcor} \\
&&=-\frac{\hbar \gamma \theta}{2 \r^3 n^3}\left(\frac{s}{n}(n_1-n_2-m_-)+m\right)
\langle{n'_1}|\frac{1}{r^3}|{n_1}\rangle \nonumber.
\eea

The energy levels (\ref{enerccm}) of the commutative Hamiltonian have a degeneracy.
It is easy to check that the multiplicity of this degeneracy at fixed $s$ reads
\beq
g^{(s)}_{n}=\sum_{|m|>[s]} (n-m_+)+\sum_{|m|\leq[s]} (n-m_+) =n^2-s^2. \label{degm}
\eeq
Thus, in order to calculate the appropriate noncommutative energy correction, we need to solve the following secular equation:
\beq
\det\left[\Delta H^{(NCM)}_{n'_1\,n_1}-\Delta E^{(NCM)}\delta_{n'_1\,n_1}\right]=0,
\eeq
where we do not take into account the dependence on $m$, by the reason of the diagonality of (\ref{matcor}) in $m$.
Here we are not going to solve this equation at least in a general sense, although at given $n$ this can be done by using computer-aided
algebraic manipulations.
We will investigate only the ground state, because, as it can be easily seen from (\ref{matcor}),
the energy correction very rapidly goes to zero when $n$ increases. Let us mention that similarly it makes no sense to investigate the case of large $s$,
because then, once again, the energy correction vanishes.

The corrections matrix (\ref{matcor}) for the ground state becomes diagonal and consequently it yields directly
the energy correction. For the ground level, at fixed $s$,
the principal quantum number takes the value $n=|s|+1$ and the azimuthal quantum number runs over the values $m=-|s|,-|s|+1,\dots,|s|-1,|s|$.
Let us remind here that the ground state of the $CCM$ has a degeneracy, in contrast to the case of the hydrogen atom.
Using the properties of the Clebsch-Gordan coefficients \cite{varsh}, for the energy correction of the ground level at fixed $s$ we get
\beq
\Delta E^{(NCM)}_{ground}=-\frac{\hbar \gamma \theta}{\r^3}\frac{m(1+|s|-s)}{(1+|s|)^5|s|(1+2|s|)}. \label{enground}
\eeq
As we can see from the above formula, the energy correction can take both negative values (for positive $m$) and positive ones (for negative $m$).
This correction is larger (in absolute value) for a negative $s$.
The degeneracy of the ground level is completely removed.
According to the general formula (\ref{degm}), the initial ground state splits into $2|s|+1$ levels.
For
fixed $s$, according to the formula (\ref{enground}), the two
extreme components of the splitted
energy levels correspond to the following
values of the azimuthal quantum number: $m=|s|$ and $m=-|s|$.
The distance between these levels is
\beq
\frac{2\hbar \gamma \theta}{\r^3}\frac{|s|(1+|s|-s)}{(1+|s|)^5|s|(1+2|s|)},
\eeq
i.e. the complete splitting of the level  is
proportional to $n^{-5}$.
The new ground level corresponds the following value of
the azimuthal quantum number: $m=|s|$.

Let us mention here that similar things can be said for higher levels.

\section{The Stark effect}

As it was mentioned above, the $CCM$ keeps most of the properties of the hydrogen atom. However, its behavior can become qualitatively
different, with regards to certain phenomena.
A representative example of this fact, is the behavior of the system in a static electric field. As it was shown in \cite{leva},
the energy correction due to the linear Stark effect has the following form:
\bea
\Delta E_{Stark}^{(CCM)}=
=\frac{3 \hbar^2 |e| \varepsilon}{2 M \gamma}\left[n\left(n_1-n_2-m_- \right)+\frac{m s}{2}\right].
\eea
As we can see, this expression differs from the formula for the usual linear Stark effect. There is an additional nontrivial correction linear in $m$,
which completely removes the degeneracy.
Thus, we can hope that the Stark effect in $NCM$ can yield results different from those obtained for the noncommutative hydrogen atom. As we know,
there is no contribution in the Stark effect due to noncommutativity \cite{chain}

Let us now calculate the Stark effect for $NCM$. For this purpose, we use the method suggested in \cite{chain}.
The potential energy of the electron in an external
electric field ${\bvep}$  is given by
\beq V_{Stark}=|e|{\bvep} \cdot {\bf r}+
\frac{|e|}{2}({\bt}\times {\bvep})\cdot\hat{\tilde{{\bf p}}}. \label{shtpot} \eeq
The change in the energy levels due to
noncommutativity (the second term in (\ref{shtpot})) is
\beq
\Delta E_{Stark}^{(NCM)}=\langle {n'_1n'_2m'}|\frac{|e|}{2}({\bt}\times
{\bvep})\cdot\hat{\tilde{{\bf p}}}|{n_1n_2m} \rangle \ . \eeq

Taking into account that $\hat{\tilde{p}}_i=\frac{M}{i\hbar}[x_i,H^{(CCM)}]$
and
$H^{(CCM)}|n_1n_2m\rangle=E^{(CCM)}_n|n_1n_2m\rangle$, the correction to the energy
levels becomes
\beq
\Delta E_{Stark}^{(NCM)}=0 ,
\eeq
meaning that the contribution to the Stark effect
due to noncommutativity vanishes. Hence, strange as it may seem, we get the same result as in the case of the noncommutative hydrogen atom.

\section{Conclusion}

In this Letter the problem of the noncommutative coulombic monopole,
i.e. the noncommutative hydrogen atom in the presence of a Dirac monopole, was investigated.
Firstly, we have constructed the corresponding Hamiltonian, starting from the fact that the turning on of noncommutativity changes only the phase space.
As a byproduct we got the relations between canonical and deformed phase variables, in the case of the arbitrary magnetic and small, but arbitrary,
"noncommutativity" fields. As it was shown here, the "noncommutative" correction yields a term which has no known analogs in the literature.

Then we discussed the energy spectrum and determined that the only "observable" effects of noncommutativity are restricted to the ground state and
possibly to the first excited level. This is a consequence of the fact that the matrix elements of the Hamiltonian correction are proportional to $1/n^k$,
where $k\ge 3$.
Therefore, we limited ourselves basically to calculating the energy correction just for the ground level.

We found that the ground state splits into $2|s|+1$ levels,
i.e. the degeneracy in the azimuthal quantum number is completely removed, as a consequence of corrections due to noncommutativity.
We can also conclude from inspecting the matrix element correction (\ref{matcor}), that the degeneracy of higher energy levels is lifted, as well,
though the exact expression of the splitted eigenvalues is of no interest, as the corresponding splitting is negligibly small.

Similarly we found that we can restrict our consideration to small values of the monopole number $s$, when we consider possibly
observable corresponding effects.
Interestingly enough, it turns out from our calculation that the absolute value of the energy correction depends on the sign of the monopole number $s$,
i.e. it is larger for negative $s$.

Finally, we studied the Stark effect for the noncommutative coulombic monopole. We have shown that, similarly to the case of the noncommutative
hydrogen atom, there are no corrections due to noncommutativity.

\subsection*{{ Acknowledgments}}

We express our gratitude to Armen Nersessian for helpful discussions. Special thanks go to Levon Mardoyan for
useful discussions and help in understanding some unclear aspects in the calculations.
This research was partially
supported by the European Community's Marie Curie Research Training Network
under contract MRTN-CT-2004-005104 Forces Universe as well as by
INTAS-00-00254 grant.


\vspace{3 mm}

\begin{thebibliography}{99}

\bibitem{string}
E.~Witten, Nucl. Phys. B {\bf 460} (1996) 33; M.~R.~Douglas and
C.~M.~Hull, J. High Energy Phys. {\bf 02} (1998) 008; C.~Chu and P.~Ho,
Nucl. Phys. B {\bf 550} (1999) 151;  Nucl. Phys. B {\bf 568}
(2000) 447; N.~ Seiberg and E.~ Witten, J. High Energy Phys. {\bf 09}
(1999) 032.

\bibitem{bran}
I.~Antoniadis, N.~Arkani-Hamed, S.~Dimopoulos and G.~R.~Dvali,
Phys.\ Lett.\ B {\bf 436} (1998) 257.

\bibitem{standart}
J.~Madore, S.~Schraml, P.~Schupp, and J.~Wess, Eur. Phys. J. C {\bf 16}
(2000) 161; X.~Calmet, B.~Jurco, P.~Schupp, J.~Wess, and
M.~Wohlgenannt, Eur. Phys. J. C {\bf 23} (2002) 363.

\bibitem{bbk}
S. Bellucci, I.L. Buchbinder and V.A. Krykhtin, Nucl. Phys. B{\bf 665} (2003) 402; ibid.
Nucl. Phys. B{\bf 693} (2004) 51.

\bibitem{hall} L.~Susskind, "The Quantum Hall Fluid and Non-Commutative Chern Simons
Theory", arXiv:hep-th/0101029; A.~P.~Polychronakos, J. High Energy
Phys. {\bf 04} (2001) 011; K.~Hasebe and Y.~Kimura, Phys.\ Lett.\ B {\bf 602} (2004) 255.

\bibitem{ncqm} J.~Gamboa, M.~Loewe and J.~C.~Rojas, Phys. Rev. D {\bf
64} (2001) 067901; B.~Morariu and A.~P.~Polychronakos, Nucl.
Phys. B {\bf 610} (2001) 531; J.~Gamboa, M.~Loewe, F.~Mendez and
J.~C.~Rojas,  Int. J. Mod. Phys. A {\bf 17} (2002) 2555.

\bibitem{ho} Pei-Ming Ho and Hsien-Chung Kao, Phys. Rev. Lett. {\bf 88} (2002)
151602.
\bibitem{scat} S.~Bellucci and A.~Yeranyan, Phys.\ Lett.\ B {\bf 609} (2005) 418.

\bibitem{magn} V.~P.~Nair and A.~P.~Polychronakos, Phys. Lett.
B{\bf 505} (2001) 267; S.~Bellucci, A.~Nersessian and C.~Sochichiu, Phys.
Lett. B{\bf 522} (2001) 345; S.~Bellucci and A.~Nersessian, Phys.
Lett. B{\bf 542} (2002) 295; S.~Bellucci, Phys. Rev. D{\bf 67} (2003) 105014.

\bibitem{zhang} J.~Z.~Zhang,  Phys.\ Rev.\ Lett.\  {\bf 93} (2004) 043002.

\bibitem{val} V.~Ter-Antonian, ``Dyon-oscillator duality,'' arXiv: quant-ph/0003106.


\bibitem{denis} D.~Khetselius, Mod.\ Phys.\ Lett.\ A {\bf 20} (2005) 263.

\bibitem{Z}
D.~Zwanziger. Phys. \ Rev. {\bf 176} (1968) 1480;
H.~McIntosh and A.~Cisneros. J.\ Math. \ Phys. {\bf 11} (1970) 896.

\bibitem{chain}
M.~Chaichian,
M.~M.~Sheikh-Jabbari and A.~Tureanu, Phys. Rev. Lett. {\bf 86}
(2001) 2716.

\bibitem{jack}
  R.~Jackiw, Int.\ J.\ Mod.\ Phys.\ A {\bf 19S1} (2004) 137.

\bibitem{mard}
 L.~Mardoyan, arXiv:quant-ph/0310143.

\bibitem{landau}
L.~D.~Landau, E.~M.~Lifshitz , Quantum mechanics: non-relativistic theory, Pergamon Press, (1977).

\bibitem{varsh}
D.~A.~Varshalovich, A.~N.~Moskalev, and V.~K.~Khersonskii. Quantum
Theory of Angular Momentum. World Scientific, Singapore, (1988).

\bibitem{leva}
 L.~Mardoyan, A.~Nersessian and M.~Petrosyan, Theor.\ Math.\ Phys.\  {\bf 140} (2004) 958, arXiv:quant-ph/0305027.


\end{thebibliography}
\end{document}